\documentclass[%
superscriptaddress,
showpacs,preprintnumbers,
nofootinbib,
 amsmath,amssymb,
 aps,
 twocolumn,
]{revtex4}
\usepackage{braket}
\usepackage{bm}
\usepackage{dcolumn}
\usepackage{cancel}
\usepackage[dvipdfmx,final]{graphicx}
\input{colordvi.tex}
\usepackage{longtable}

\def\bea{\begin{eqnarray}}
\def\eea{\end{eqnarray}}

\newcommand\unit[1]{\,{\rm #1}}

\newcommand\keV{\unit{keV}}
\newcommand\MeV{\unit{MeV}}
\newcommand\GeV{\unit{GeV}}
\newcommand\TeV{\unit{TeV}}

\usepackage{color}
\usepackage{ulem}

\begin{document}
\preprint{CTPU-PTC-19-01 RESCEU-1/19}
\title{Smallest Halos in Thermal Wino Dark Matter}

\author{Shin'ichiro Ando}
\email{s.ando@uva.nl}
\affiliation{GRAPPA Institute, University of Amsterdam, 1098 XH Amsterdam, Netherlands}
\affiliation{Kavli Institute for the Physics and Mathematics of the Universe (WPI), University of Tokyo, Kashiwa, Chiba 277-8583, Japan}
\author{Ayuki Kamada}
\email{akamada@ibs.re.kr}
\affiliation{Center for Theoretical Physics of the Universe, Institute for Basic Science (IBS), Daejeon 34126, Korea}
\author{Toyokazu Sekiguchi}
\email{sekiguti@resceu.s.u-tokyo.ac.jp}
\affiliation{Research Center for the Early Universe (RESCEU), Graduate School of Science, The University of Tokyo, Tokyo 113-0033, Japan}
\affiliation{Institute of Particle and Nuclear Studies, High Energy Accelerator Research Organization (KEK), Oho 1-1, Tsukuba 305-0801, Japan}
\author{Tomo Takahashi}
\email{tomot@cc.saga-u.ac.jp}
\affiliation{Department of Physics, Saga University, Saga 840-8502, Japan
}

\date{\today}

\begin{abstract}
(Mini) split supersymmetry explains the observed Higgs mass and evades stringent constraints, while keeping good features of TeV-scale supersymmetry other than the little hierarchy problem.
Such scenarios naturally predict thermal wino dark matter whose mass is around $3 \TeV$.
Its non-perturbatively enhanced annihilation is a promising target of indirect detection experiments.
It is known that identifying the smallest halos is essential for reducing an uncertainty in interpreting indirect detection experiments.
Despite its importance, the smallest halos of thermal wino dark matter have not been well understood and thus are investigated in this work.
In particular, we remark on two aspects:
1) the neutral wino is in kinetic equilibrium with primordial plasma predominantly through inelastic processes involving the slightly heavier charged wino;
and 2) the resultant density contrast shows larger powers at dark acoustic oscillation peaks than in cold dark matter, which is known as an overshooting phenomenon.
By taking them into account, we provide a rigorous estimate of the boost factor.
Our result facilitates accurately pinning down thermal wino dark matter through vigorous efforts in indirect detection experiments.
\end{abstract}


\maketitle

\section{Introduction}
The identity of dark matter and the origin of electroweak symmetry breaking are big mysteries of modern particle physics.
A prominent possibility of addressing these two issues is supersymmetry.
Supersymmetry controls a scalar potential by relating the scalar with a partner fermion.
Electroweak symmetry breaking is insensitive to higher energy physics through the non-renormalization theorem and is driven predominantly by supersymmetry breaking.
Furthermore, supersymmetry provides the lightest supersymmetric particle as a dark matter candidate.

In particular, (mini) split supersymmetry~\cite{ArkaniHamed:2004fb, Giudice:2004tc, Wells:2004di} (see also Refs.~\cite{Ibe:2011aa, Ibe:2012hu, Arvanitaki:2012ps, Hall:2012zp, ArkaniHamed:2012gw}), where scalar (other than the Higgs boson) masses are ${\cal O}(100\text{--}1000) \TeV$ and fermion masses are ${\cal O}(0.1\text{--}1) \TeV$, attracts growing interest.
Heavy scalars and gravitino evade the constraints from collider searches, flavor physics, and cosmological problems, which TeV-scale supersymmetry suffers from, although supersymmetry breaking scale is mildly far from the electroweak scale (little hierarchy problem).
Light gauginos drive a precise grand unification~\cite{Hisano:2013cqa, Hisano:2013exa} and also provide experimental windows on this framework.
The anomaly mediation contribution~\cite{Randall:1998uk, Giudice:1998xp} is a promising dominant source of gaugino masses, where the wino is the lightest supersymmetric particle.
In the following, we consider pure wino-like dark matter, where the higgsino is as heavy as scalars and the gauge interaction dominates the interaction of the wino.
The neutral wino $\chi^{0}$ is accompanied by the slightly heavier charged wino $\chi^{\pm}$.
The mass difference is dominated by a loop contribution, $\Delta m_{\chi} = m_{\chi^{\pm}} - m_{\chi^{0}} \simeq 160 \text{--} 170 \MeV$~\cite{Cheng:1998hc, Feng:1999fu, Gherghetta:1999sw, Ibe:2012sx}.
The charged wino is thus long lived and leaves significant signals such as a disappearing track in projected high-energy colliders~\cite{Low:2014cba, Cirelli:2014dsa, diCortona:2014yua, Mahbubani:2017gjh, Fukuda:2017jmk, Han:2018wus, Saito:2019rtg}.

The wino thermal relic explains the observed dark matter abundance when $m_{\chi} \simeq 2.7 \text{--} 3.0 \TeV$~\cite{Hisano:2006nn, Hryczuk:2010zi, Beneke:2014hja, Beneke:2016ync, Mitridate:2017izz}.
Its annihilation cross section is enhanced in the present Universe by a non-perturbative effect, known as Sommerfeld enhancement~\cite{Hisano:2002fk, Hisano:2003ec, Hisano:2004ds, Hisano:2005ec}.
This is why thermal wino dark matter has been and is encouraged to be intensively searched in indirect detection experiments~\cite{Ullio:2001qk, Chattopadhyay:2006xb, Grajek:2008jb, Belanger:2012ta, Cohen:2013ama, Fan:2013faa, Hryczuk:2014hpa, Bhattacherjee:2014dya, Cirelli:2015bda, Shirasaki:2016kol, Lefranc:2016fgn} as a clue to split supersymmetry.
To lay siege to wino dark matter by accumulating all these efforts, reducing
theoretical uncertainties is crucial.
For example, infrared divergences and associated resummation for the wino annihilation cross
section to electroweak gauge bosons have been studied intensively~\cite{Hryczuk:2011vi, Baumgart:2014vma, Ovanesyan:2014fwa, Baumgart:2014saa, Baumgart:2015bpa, Baumgart:2017nsr, Rinchiuso:2018ajn, Baumgart:2018yed}.

This article is devoted to providing better knowledge on the smallest halos of thermal wino dark matter.
It would play an essential role in reducing a theoretical uncertainty arising from enhancement of the annihilation rate through dark matter clumping~\cite{Silk:1992bh, Bergstrom:1998jj, Baltz:1999ra, Bergstrom:2000bk}.
This enhancement is called a flux multiplier and is very significant especially in extragalactic gamma-ray searches~\cite{Bergstrom:2001jj, Ullio:2002pj, Taylor:2002zd, Elsaesser:2004ap, Elsaesser:2004ck, Ando:2005hr, Ando:2005xg}.
Once a cross-correlation with large-scale structure of the Universe is taken as recently proposed~\cite{Camera:2012cj, Ando:2013xwa, Fornengo:2013rga, Shirasaki:2014noa, Ando:2014aoa, Fornengo:2014cya, Camera:2014rja, Xia:2015wka}, the clumping uncertainty is reduced to a boost factor.
We are encouraged by its promising potential to pin down wino dark matter~\cite{Shirasaki:2016kol}, since the cross-correlation with large-scale structure will be statistically improved in near future wide-field surveys.

There have been attempts to determine the smallest halos in supersymmetric dark matter~\cite{Gondolo:2004sc, Profumo:2006bv, Bringmann:2006mu, Bringmann:2009vf, Cornell:2012tb, Bringmann:2018lay} and even particularly in wino dark matter~\cite{Diamanti:2015kma}.
However, previous literatures did not appreciate the fact that what keeps wino dark matter in kinetic equilibrium with primordial plasma is not elastic processes but inelastic ones unlike typical bino dark matter.
Indeed it is known that the wino-nucleon scattering cross section is suppressed in the decoupling limit of the higgsino, while loop contributions barely keep the cross section just above the neutrino background in direct detection experiments~\cite{Hisano:2004pv, Hisano:2010fy, Hisano:2010ct, Hisano:2011cs, Hisano:2012wm, Cheung:2014hya, Hisano:2015rsa}.
This subtlety on wino dark matter kinetic decoupling has been studied for the temperature evolution in Ref.~\cite{Arcadi:2011ev}, but not yet for the evolution of the primordial density contrast, which determines dark matter clumping in the present Universe.

As we see, the fact that wino dark matter is kinetically equilibrated through an inelastic process results in an enhanced oscillation of the matter power spectrum, which originates from ``overshooting"~\cite{Kamada:2016qjo,Sarkar:2017vls,Kamada:2017oxi}; namely, dark acoustic oscillation peaks have a larger power than the density contrast of cold dark matter.
This overshooting phenomenon was discovered for the first time in the study of electromagnetically charged dark matter in Ref.~\cite{Kamada:2016qjo} and confirmed by Ref.~\cite{Sarkar:2017vls}.  Its underlying physics was clarified in Ref.~\cite{Kamada:2017oxi}.
Our present work demonstrates that overshooting of the matter power spectrum can be seen broadly in minimal
dark matter~\cite{Cirelli:2005uq, Cirelli:2007xd}, although we focus on wino dark matter.

\section{Relevant Processes}
Elastic scattering of the neutral wino in late-time thermal bath, long after the wino freezes out, is suppressed in the decoupling limit of the higgsino~\cite{Hisano:2000dz, Ibe:2012hr}.
The leading contribution arises from a one-loop diagram with $W$-boson exchange~\cite{Ibe:2012hr}.
The collision term for the neutral wino phase space distribution $f_{\chi^{0}}$ can be approximated by the Fokker-Planck form~\cite{Binder:2016pnr},
\begin{eqnarray}
\label{eq:elacoll}
\frac{1}{E} C_{\chi^{0}, {\rm ela}} \approx g_{\chi^{0}} \gamma_{\rm ela} \frac{\partial}{\partial {\bf p}} \cdot 
\left[ m_{\chi} T \frac{\partial}{\partial {\bf p}} f_{\chi^{0}} + ({\bf p} - m_{\chi} {\bf u}) f_{\chi^{0}} \right] \,,
\end{eqnarray}
with the three momentum of the wino ${\bf p}$, the temperature $T$, and the bulk motion ${\bf u}$ of the thermal bath.  $g_{\chi^{0}} = 2$ is neutral wino internal degrees of freedom.
The momentum transfer rate $\gamma_{\rm ela}$ is given by
\begin{eqnarray}
\label{eq:gamela}
\gamma_{\rm ela} = 8 \frac{100}{\pi^{3}} g_{\rm loop}^{2} G_{F}^{4} m_{W}^{4} \frac{T^{6}}{m_{\chi}} \,,
\end{eqnarray}
with the Fermi constant $G_{F} \simeq 1.2 \times 10^{-5} \GeV^{-2}$, the $W$-boson mass $m_{W} \simeq 80 \GeV$, and
\begin{eqnarray}
g_{\rm loop} &=& \frac{1}{3 \pi^{2}} \left( 2 (8 - \omega - \omega^{2}) \sqrt{\frac{\omega}{4 - \omega}}  \arctan \left( \sqrt{\frac{4 - \omega}{\omega}} \right) \right. \notag \\
&& \left.  - \omega \left( 2 - (3 + \omega) \ln \omega \right) \right) \,,
\end{eqnarray}
with $\omega = m_{W}^{2} / m_{\chi}^{2}$.
As we see, this elastic process is subdominant when compared to inelastic processes in keeping the neutral wino in kinetic equilibrium with the heat bath.

A key observation is that the charged wino is in kinetic equilibrium with the heat bath through efficient electromagnetic interactions;
and thus its phase space distribution follows
\begin{eqnarray}
f_{\chi^{\pm}} ({\bf p}) \approx \frac{n_{\chi^{\pm}}}{g_{\chi^{\pm}}} \left( \frac{2 \pi}{m_{\chi} T} \right)^{3 / 2} \exp \left( - \frac{({\bf p} - m_{\chi} {\bf u})^{2}}{2 m_{\chi} T} \right) \,,
\end{eqnarray} 
where $g_{\chi^{\pm}} = 4$ is charged wino internal degrees of freedom.
Resultantly, inelastic processes between the neutral wino and charged wino can keep the neutral wino in kinetic equilibrium with the heat bath.

Furthermore, a kick momentum through the mass deficit in an inelastic process is negligible when compared to a typical wino momentum $\sim \sqrt{m_{\chi} T}$ until a very late time, $T \sim 10 \keV$, since the wino mass difference $\Delta m_{\chi} \simeq 160 \MeV$ is much smaller than the wino mass $m_{\chi} \simeq 3 \TeV$. 
The collision term can be approximated by a pure conversion from~\cite{Arcadi:2011ev},
\begin{eqnarray}
\label{eq:inelacoll}
\frac{1}{E} C_{\chi^{0}, {\rm inela}} &\approx& g_{\chi^{\pm}} ( \Gamma_{\rm dec} + \Gamma_{\rm inela} ) \left( f_{\chi^{\pm}} - f_{\chi^{0}} e^{- \Delta m_{\chi} / T} \right) \,. \notag \\
\end{eqnarray}
There are two contributions: decay, 
\begin{eqnarray}
\label{eq:gamdec}
\Gamma_{\rm dec} \approx \frac{f_{\pi}^{2} G_{F}^{2} |V_{u d}|^{2}}{\pi} \Delta m_{\chi}^{3} \sqrt{1 - \frac{m_{\pi^{\pm}}^{2}}{\Delta m_{\chi}^{2}}} \,,
\end{eqnarray}
where $f_{\pi} \simeq 130 \MeV$ is the pion decay constant, $m_{\pi^{\pm}} \simeq 140 \MeV$ is the charged pion mass, and $|V_{u d}| \simeq 0.97$ is the first generation diagonal component of the CKM matrix; and inelastic scattering,
\begin{eqnarray}
\label{eq:gaminela}
\Gamma_{\rm inela} \approx 2 \frac{8 G_{F}^{2}}{\pi^{3}} T^{3} \left( \Delta m_{\chi}^{2} + 6 \Delta m_{\chi} T + 12 T^{2} \right) \,.
\end{eqnarray} 

We remark that the inelastic reaction rate for the charged wino, $\Gamma_{\rm dec} + \Gamma_{\rm inela}$, is much larger than the Hubble expansion rate $H$.
Resultantly, the charged wino is in chemical equilibrium with the neutral wino.
Wino number densities,
\begin{eqnarray}
n_{\chi^{\pm}} = g_{\chi^{\pm}} \int \frac{d^{3} {\bf p}}{(2 \pi)^{3}} f_{\chi^{\pm}} \,, \quad  n_{\chi^{0}} = g_{\chi^{0}} \int \frac{d^{3} {\bf p}}{(2 \pi)^{3}} f_{\chi^{0}} \,,
\end{eqnarray}
satisfy
\begin{eqnarray}
\label{eq:nchipm}
&& n_{\chi^{\pm}} \approx \frac{g_{\chi^{\pm}}}{g_{\chi^{0}}} n_{\chi^{0}} e^{-\Delta m_{\chi}/T} \,, \\
&& {\dot n}_{\chi^{0}} + 3 \frac{\dot a}{a} n_{\chi^{0}} = - \left( {\dot n}_{\chi^{\pm}} + 3 \frac{\dot a}{a} n_{\chi^{\pm}} \right) \approx 0 \,.
\end{eqnarray}
The last equality is valid when $T \ll \Delta m_{\chi}$.
Here a dot denotes a derivative with respect to the conformal time.
We use the synchronous gauge following the notation of Ref.~\cite{Ma:1995ey},
\begin{eqnarray}
ds^{2} = a^{2} (- d\tau^{2} + (\delta_{i j} + h_{i j}) dx^{i} dx^{j}) \,,
\end{eqnarray}
with $h_{i j}$ being the metric perturbation in the Fourier space ${\bf k} = |{\bf k}| {\hat {\bf k}}$ and decomposed as
\begin{eqnarray}
h_{i j} = {\hat k}_{i} {\hat k}_{j} h + \left( {\hat k}_{i} {\hat k}_{j} - \frac{1}{3} \delta_{i j} \right) \eta \,.
\end{eqnarray}
Until the next section, we consider only homogeneous and isotropic components.

\begin{figure}[!h]
\centering
\includegraphics[scale=0.6]{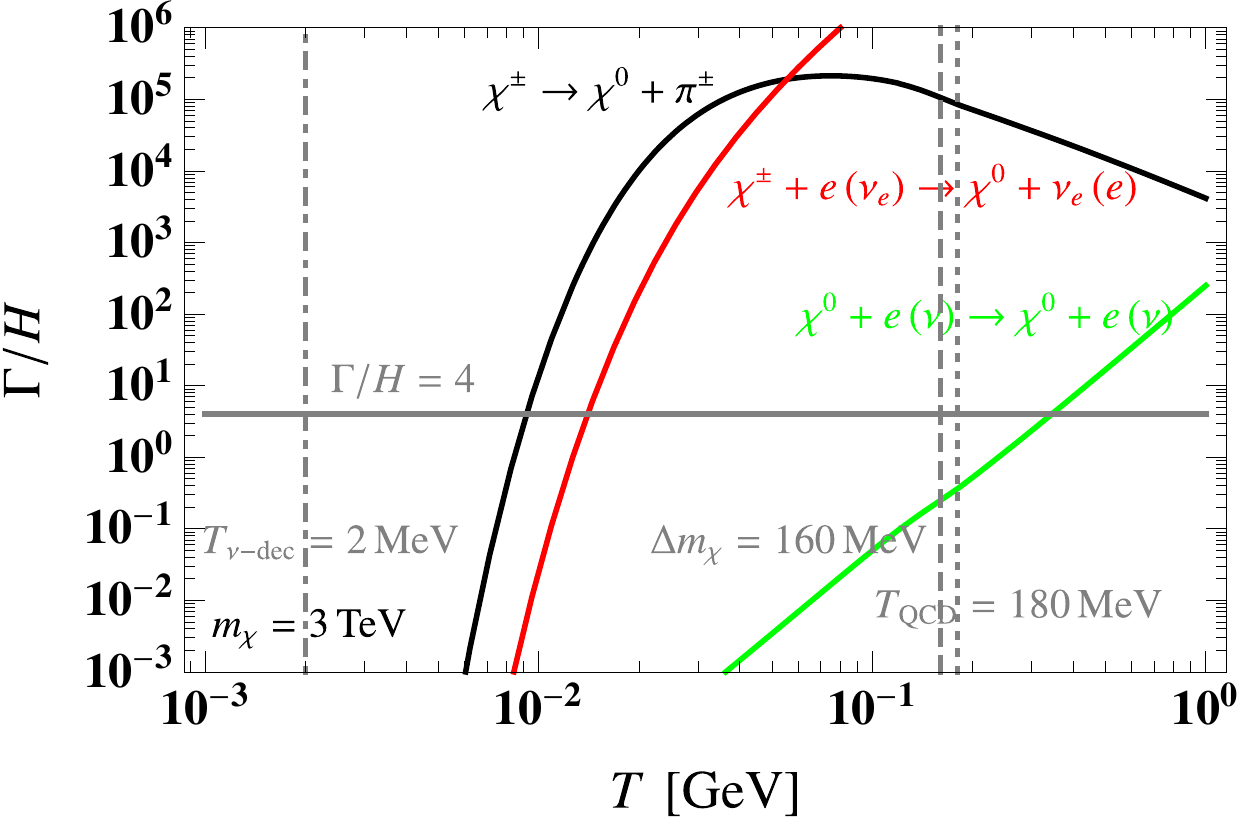}
\caption{Reaction rates in the temperature evolution equation, $g_{\chi^{\pm}} \Gamma_{\rm dec} \exp \left( - \Delta m_{\chi} / T \right)$ (black), $g_{\chi^{\pm}} \Gamma_{\rm inela} \exp \left( - \Delta m_{\chi} / T \right)$ (red), and $2 g_{\chi^{0}} \gamma_{\rm ela}$ (green), normalized by the Hubble expansion rate $H$.}
\label{fig:intrateT}
\end{figure}

The evolution of the neutral wino temperature,
\begin{eqnarray}
3 m_{\chi} T_{\chi^{0}} n_{\chi^{0}} &=& g_{\chi^{0}} \int \frac{d^{3} {\bf p}}{(2 \pi)^{3}} {\bf p}^{2} f_{\chi^{0}}  \,,
\end{eqnarray}
is governed by
\begin{eqnarray}
{\dot T}_{\chi^{0}} + 2 \frac{\dot a}{a} T_{\chi^{0}} &\approx& a \Bigg[ g_{\chi^{\pm}} ( \Gamma_{\rm dec} + \Gamma_{\rm inela} ) e^{-\Delta m_{\chi}/T} \notag \\
&&  + 2 g_{\chi^{0}} \gamma_{\rm ela} \Bigg] \left( T - T_{\chi^{0}} \right) \,.
\end{eqnarray}
Figure~\ref{fig:intrateT} compares the elastic and inelastic reaction rates in this equation.
As one can see, the elastic scattering decouples earlier than the inelastic processes.
Thus we neglect the elastic scattering in the following.
The inelastic reaction rates decrease rapidly with decreasing temperature below $T \simeq \Delta m_{\chi}$ due to the Boltzmann suppression, and drops below the Hubble expansion rate around $T \simeq 9.2 \MeV$.

Until then, the neutral wino is in kinetic equilibrium with primordial plasma and thus is involved in an acoustic oscillation driven by photon pressure, called a dark acoustic oscillation.
Na\"{i}vely, growth of the density contrast is suppressed below the horizon scale corresponding to kinetic decoupling.
We define the epoch of kinetic decoupling with $(\Gamma / H) |_{T =  T_{\rm kd}} = 4$ as done in Ref.~\cite{Bertschinger:2006nq}.
One could identify $T \simeq 9.2 \MeV$ as a kinetic decoupling temperature, at which the horizon scale is  $k_{\rm kd} = 1 / \tau_{\rm kd} \simeq 0.11 \unit{/pc}$, and estimate the smallest protohalo mass as
\begin{eqnarray}
\label{eq:Mnaive}
M_{\rm kd}^{\text{na\"{i}ve}} &=& \frac{4 \pi}{3} \rho_{\chi, 0} \tau_{\rm kd}^{3} \notag \\
&\simeq& 1.1 \times 10^{-4} \, M_{\odot} \left( \frac{T_{\rm kd}}{9.2 \MeV} \sqrt{\frac{g_{\rm eff}}{10.75}} \right)^{-3} \,, \quad
\end{eqnarray}
where $g_{\rm eff}$ counts the effective massless degrees of freedom.
Here we replace the present total matter mass density $\rho_{m, 0}$ in Ref.~\cite{Bertschinger:2006nq} by the present dark matter mass density $\rho_{\chi, 0}$ since the baryon does not collapse into such a small halo even with cold dark matter due to gas pressure.
Our na\"{i}ve estimate is already different from an estimate in the previous literature~\cite{Diamanti:2015kma} only considering elastic processes, where $T_{\rm kd} \sim 1 \GeV$ and $M_{\rm kd} \sim 10^{-11} \, M_{\odot}$.
However, as we see in the next section, the true evolution of the density contrast is more complicated than
na\"{i}vely expected due to the overshooting phenomenon.

\section{Evolution of Density Perturbations}
We remark again that the charged wino is in chemical equilibrium with the neutral wino so that the evolution of wino (dimensionless) density perturbations,
\begin{eqnarray}
&& n_{\chi^{\pm}} \delta_{\chi^{\pm}} = g_{\chi^{\pm}} \int \frac{d^{3} {\bf p}}{(2 \pi)^{3}} \, \delta f_{\chi^{\pm}} \,, \\
&& n_{\chi^{0}} \delta_{\chi^{0}} = g_{\chi^{0}} \int \frac{d^{3} {\bf p}}{(2 \pi)^{3}} \, \delta f_{\chi^{0}} \,,
\end{eqnarray}
satisfies
\begin{eqnarray}
\label{eq:deltachi}
&&\delta_{\chi^{\pm}} \approx \delta_{\chi^{0}} + \frac{\Delta m_{\chi}}{T} \delta_{T} \,, \\
&&{\dot \delta}_{\chi^{0}} + \theta_{\chi^{0}} + \frac{1}{2} {\dot h} \notag \\
&& = -\left(\frac{\dot n_{\chi^0}}{n_{\chi^0}}+3\frac{\dot a}{a}\right)(\delta_{\chi^0}-\delta_{\chi^\pm}) - \frac{n_{\chi^{\pm}}}{n_{\chi^{0}}} \left( {\dot \delta}_{\chi^{\pm}} + \theta_{T} + \frac{1}{2} {\dot h} \right) \notag \\
&& \approx 0 \,,
\end{eqnarray}
where $\delta_{T} = \delta T / T$ and $\theta_{T} = i {\bf k} \cdot {\bf u}$ are (dimensionless) temperature perturbation and velocity potential of the thermal bath, respectively.
The last equality is valid when $T \ll \Delta m_{\chi}$.

Meanwhile the evolution of velocity potential of the neutral wino,
\begin{eqnarray}
m_{\chi^{0}} n_{\chi^{0}} \theta_{\chi^{0}} &=& g_{\chi^{0}} \int \frac{d^{3} {\bf p}}{(2 \pi)^{3}} (i {\bf k} \cdot {\bf p}) \delta f_{\chi^{0}} \,,
\end{eqnarray}
is governed by
\begin{eqnarray}
\label{eq:thetachi}
{\dot \theta}_{\chi^{0}} + \frac{\dot a}{a} \theta_{\chi^{0}} \approx a g_{\chi^{\pm}} ( \Gamma_{\rm dec} + \Gamma_{\rm inela} )  e^{-\Delta m_{\chi}/T} \left( \theta_{T} - \theta_{\chi^{0}} \right) \,.
\notag \\
\end{eqnarray}
Here we ignore sound speed of the neutral wino.
We take into account the sound speed or generically free-streaming effect by multiplying the resultant $\delta_{\chi^{0}}$ by
\begin{eqnarray}
\exp \left( - \frac{k^{2}}{2 k_{\rm fs}(\tau)^{2}} \right) \,,
\end{eqnarray}
with
\begin{eqnarray}
k_{\rm fs}^{-1}
&=& \sqrt{ \frac{6 T_{\rm kd}}{5 m_{\chi^{0}}} } \int_{\tau_{*}}^{\tau} \frac{d \tau'}{a (\tau') / a_{\rm kd}} \notag \\
&\approx& \sqrt{ \frac{6 T_{\rm kd}}{5 m_{\chi^{0}}} } \tau_{\rm kd} \ln \left( \frac{\tau_{\rm eq}}{\tau_{*}} \right) \,,
\end{eqnarray}
and $\tau_{*} = 1.05 \tau_{\rm kd}$, as suggested in Ref.~\cite{Bertschinger:2006nq}.
The second equality is valid long after matter radiation equality and $\tau_{\rm eq}$ is the conformal
time at matter radiation equality.

\begin{figure}[!h]
\centering
\includegraphics[scale=0.6]{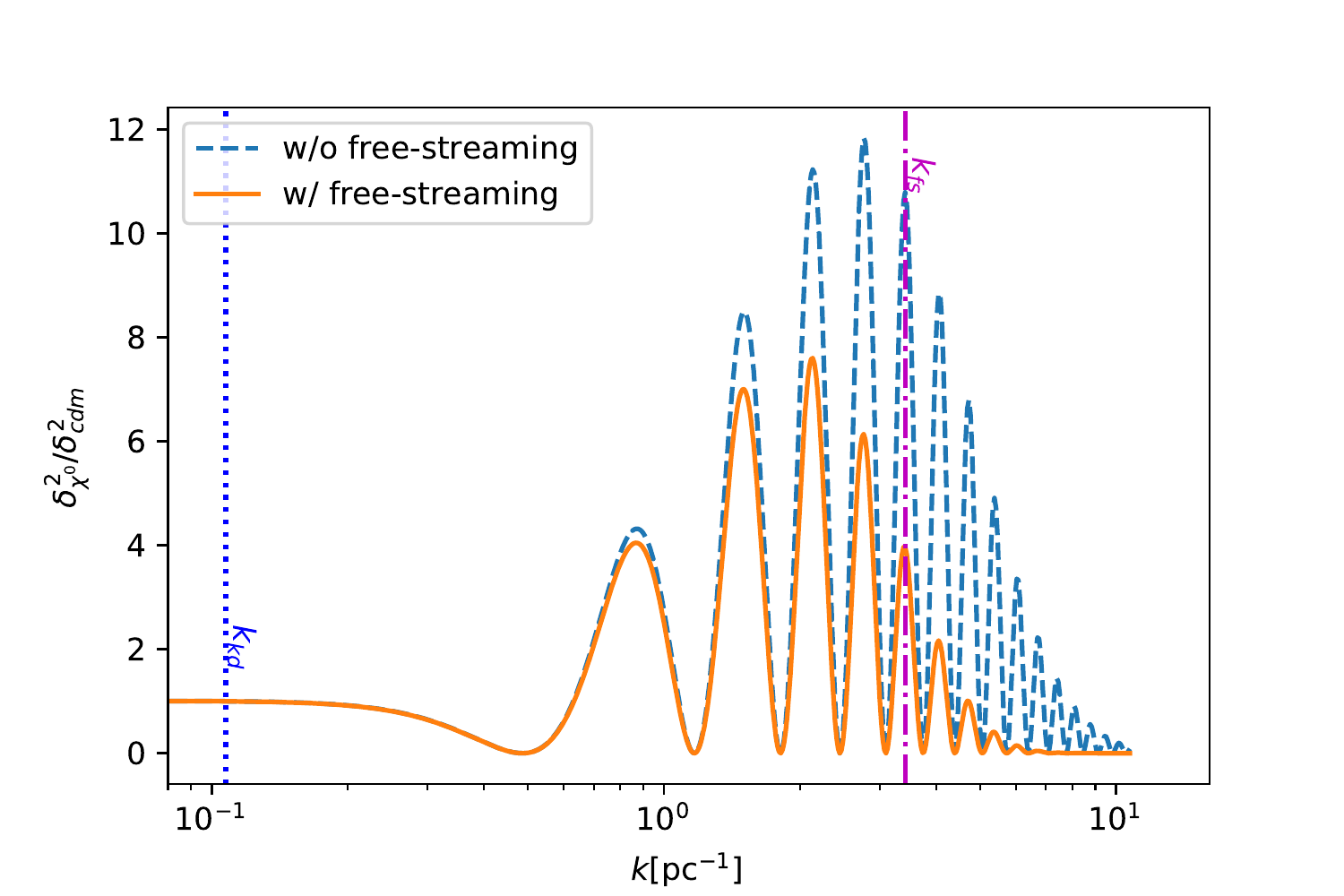}
\caption{Wino dark matter power spectrum at $\tau_{\rm eq}$ normalized to the cold dark matter one.}
\label{fig:ratio}
\end{figure}

One obtains a closed set of equations by combining the above wino equations with the radiation equations and the Einstein equations~\cite{Ma:1995ey}.
We start our numerical integration of $\delta_{\chi}$ when the mode is in superhorizon and the neutral wino tightly couples with radiation.
We stop it when the neutral wino kinetically decouples and $\delta_{\chi}$ starts to logarithmically grow in deep
subhorizon as $\delta_{c}$.
Figure~\ref{fig:ratio} shows the resultant wino dark matter power spectrum at $\tau_{\rm eq}$ normalized to the cold dark matter one.
One can see a dark acoustic oscillation below the horizon scale at kinetic decoupling as expected.
A striking feature is that peak powers of the dark acoustic oscillation are $\sim 10$ times larger than the cold dark matter powers.
This is the overshooting phenomenon, which takes place when kinetic decoupling proceeds suddenly, ${\dot \gamma} / H \gg a H$ at $\gamma / H = 4$~\cite{Kamada:2017oxi}.

As is well known in the baryon acoustic oscillation~\cite{Hu:1994uz, Hu:1994jd, Hu:1995en}, the oscillation amplitude of an acoustic wave is constant and undamped as long as radiation dominates the entropy of primordial plasma.
Damping of a dark acoustic oscillation arises from intermittent collisions around kinetic decoupling, which mixes up different oscillation phases and averages out the oscillation.
This is why it is sometimes called Landau damping in the literature.
If kinetic decoupling proceeds instantaneously, there is no damping mechanism other than free-streaming of a dark matter particle or a particle with which a dark matter particle scatters (collisionless damping or Silk damping~\cite{Silk:1967kq}, respectively).
Reference~\cite{Kamada:2017oxi} argues that in such a case, not only is the dark acoustic oscillation undamped, but its peak powers also exceed cold dark matter powers. 
This is because a supersonic motion of dark matter fluid, which is a remnant of the dark acoustic oscillation, further compresses dark matter fluid after kinetic decoupling.

In Fig.~\ref{fig:ratio}, the wino matter power spectrum multiplied by the free-streaming factor is also shown.
One sees that damping of dark acoustic oscillation peaks is determined practically by free-streaming, $k_{\rm fs} \simeq 3.5 \unit{/pc}$ for $T_{\rm kd} = 9.2 \MeV$.
Its cutoff mass is estimated by analogy to the Jeans mass as~\cite{Green:2005fa}
\begin{eqnarray}
M_{\rm fs} &=& \frac{4 \pi}{3} \rho_{\chi, 0} \left( \frac{\pi}{k_{\rm fs}} \right)^{3} \notag \\
&\simeq& 1.0 \times 10^{-7} \, M_{\odot} \left( \frac{T_{\rm kd}}{9.2 \MeV} \sqrt{\frac{g_{\rm eff}}{10.75}} ~\right)^{-3} \notag \\
&& \times \left( \ln \left( \frac{T_{\rm kd}}{9.2 \MeV} \sqrt{\frac{g_{\rm eff}}{10.75}}  ~\right) \right)^{3} \,.
\end{eqnarray}
We note that fudge factors, $(\gamma / H)|_{T = T_{\rm kd}} = 4$ and $\tau_{*} / \tau_{\rm kd} = 1.05$, are calibrated in bino dark matter~\cite{Bertschinger:2006nq}.
These values could change only slightly in our wino dark matter case.
One needs to follow a full Boltzmann hierarchy to identify the precise values, but it is beyond the scope of this work.
One also needs to take into account quantum chromodynamics phase transition, neutrino decoupling, and $e^{\pm}$ annihilation.
We estimate that they may change our result up to 10\%.
(This estimation is presented in Appendix~\ref{sec:evoldelta}.)

\section{Impacts on Indirect Detections} 
To demonstrate the impact of the matter power spectrum in wino dark matter on indirect detection experiments, we compute the annihilation boost factor $B$, which 
is defined for a given field halo mass $M$ as
\begin{eqnarray}
L (M) = (1 + B (M)) {\bar L} (M) \,,
\end{eqnarray}
where $L$ is the total luminosity and ${\bar L}$ is the luminosity of the smooth component. $B(M)$ is a sum of subhalo contributions~\cite{Strigari:2006rd, Kuhlen:2008aw, Ando:2019xlm},
\begin{eqnarray}\label{eq:boost}
B (M) = \frac{1}{{\bar L} (M)} \int^{M}_{m_{\rm min}} dm \frac{dN_{\rm sh}}{dm}{ L}_{\rm sh} (m) \,,
\end{eqnarray}
with $dN_{\rm sh} / dm$ being the subhalo mass function.
To compute Eq.~\eqref{eq:boost}, one needs to know $dN_{\rm sh} / dm$ as well as the density profile of subhalos.
We follow the method of Ref.~\cite{Hiroshima:2018kfv}, which analytically describes how these quantities evolve.
They successfully reproduce results of N-body simulations, in particular, by taking into account tidal mass stripping that subhalos undergo inside their hosts. 
(A brief summary of the method is presented in Appendix~\ref{sec:calcB}.)

Figure~\ref{fig:Bsh} shows the subhalo boost factor $B(M)$ of dark matter annihilation as a function of the mass of host halos $M$.
It is manifested that, in the case of wino dark matter, the substructure boost factor is significantly enhanced when compared to the case of the na\"{i}ve model featuring a sudden cutoff at $M_{\rm fs} = 10^{-7}M_\odot$ (which we refer to as CDM).
We find that the effect is as large as $\sim$30\% for relatively large halos (galaxies, clusters, etc.), even though this affects only subhalos with very small masses, $m \lesssim 10^{-5} M_{\odot}$.

\begin{figure}[!h]
\centering
\includegraphics[width=8.5cm]{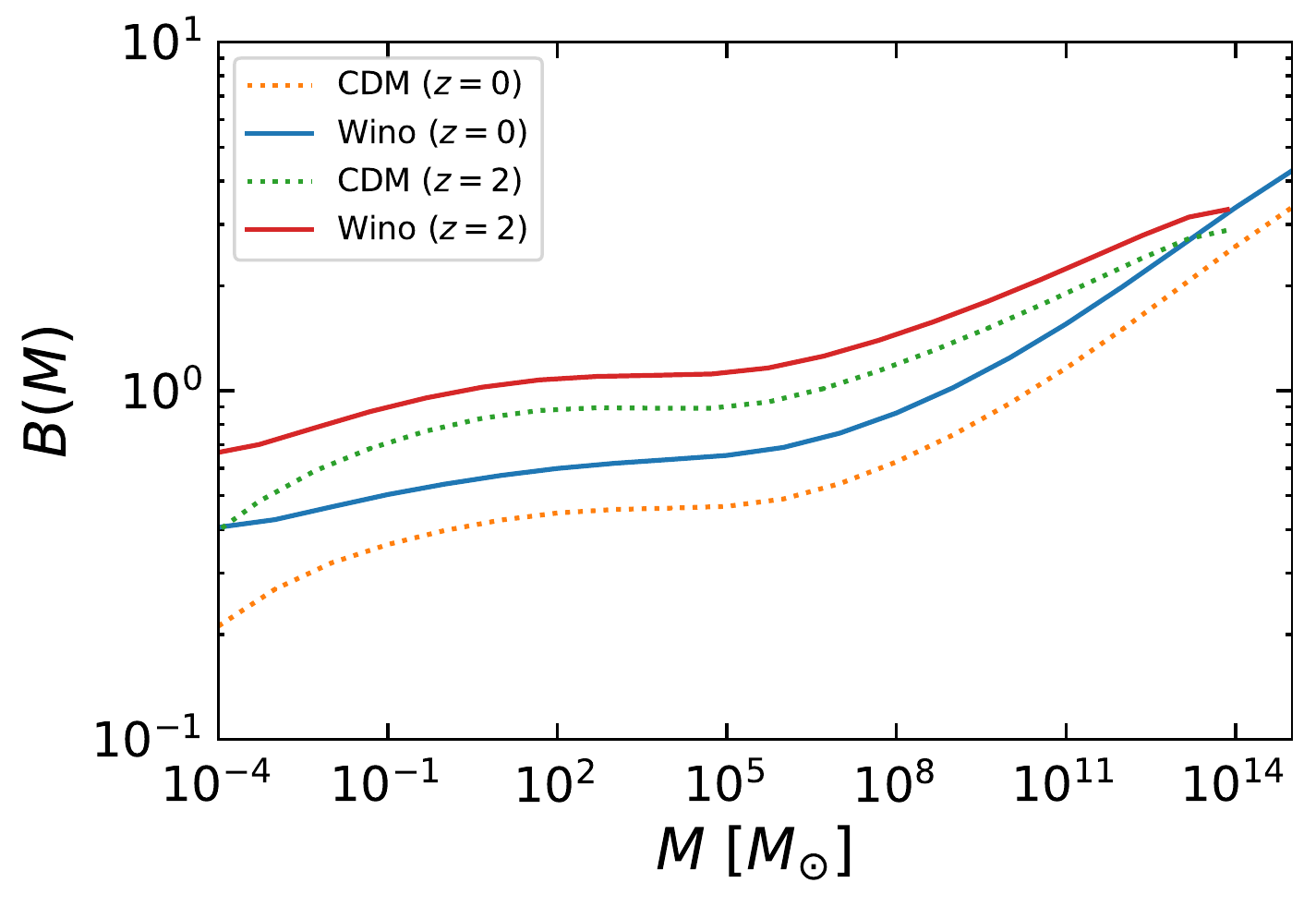}
\caption{The annihilation boost factor $B(M)$ as a function of the mass of host halos $M$ at $z=0$ and 2. The case of wino dark matter (solid) is compared with that of  CDM (dotted).}
\label{fig:Bsh}
\end{figure}

\section{Concluding Remarks} 
Thermal wino dark matter is a promising clue of split supersymmetry, which can be considered as  one of the most attractive new physics after the Higgs discovery.
A combination of indirect detection experiments has the potential to rigorously explore wino dark matter in the near future.
To this end, astrophysical uncertainties should be understood in more detail.
In this work, we have provided a rigorous estimate of the boost factor by taking into account two overlooked aspects of the wino dark matter density contrast.
First, the neutral wino is in kinetic equilibrium with primordial plasma not through elastic processes, but through inelastic processes involving the charged wino.
Resultantly, kinetic decoupling temperature is around $9.2 \MeV$, 2 orders of magnitude smaller than expected in the previous literature.
Second, a dark acoustic oscillation of the neutral wino shows the overshooting phenomenon; namely, its peak powers of dark acoustic oscillations are larger than the cold dark matter case.
It follows that free-streaming after kinetic decoupling, rather than the dark acoustic oscillation, determines the smallest halos of wino dark matter.
The implications of our result are not limited in indirect detection experiments.
Potential investigation of the small-size halo abundance, e.g., in a pulsar timing array~\cite{Ishiyama:2010es, Kashiyama:2018gsh}, has been proposed.
Further studies including a set of dedicated $N$-body simulations are warranted.

\acknowledgements
A. K. thanks Satoshi Shirai, Masato Shirasaki, and Naoki Yoshida for encouraging A. K. to work on this project and providing valuable comments on the manuscript.
T. S. thanks Piero Ullio for discussion.  
This work is supported by Institute for Basic Science under the project code, IBS-R018-D1 (A. K.), JSPS KAKENHI Grant Numbers JP17H04836 (S. A.), 15K05084 (T. T.), 17H01131 (T. T.), JP15H02082 (T. S.), 18H04339 (T. S.), 18K03640 (T. S.), and MEXT KAKENHI Grant Number 15H05888 (T. T.), JP18H04340 (S. A.), JP18H04578 (S. A.).  \\


\appendix




\section{Evolution of Density Perturbations}
\label{sec:evoldelta}
First, we summarize 1) the radiation equations and 2) the Einstein equations, which are followed simultaneously with the wino equations [Eqs.~\eqref{eq:deltachi} and \eqref{eq:thetachi}].

1) Before neutrino decoupling around $T \sim 2 \MeV$, one can reliably use a perfect fluid description of primordial plasma,
\begin{eqnarray}
\label{eq:cont_rad}
&& {\dot \delta}_{r} + \frac{4}{3} \theta_{r} + \frac{2}{3} {\dot h} = 0 \,, \\
\label{eq:eular_rad}
&& {\dot \theta}_{r} - \frac{1}{4} \delta_{r} = 0 \,, \\
\label{eq:sigma_rad}
&& \sigma_{r} = 0 \,,
\end{eqnarray}
where $\delta_{r}$, $\theta_{r}$, and $\sigma_{r}$ are density perturbation, velocity potential, and anisotropic stress of radiation, respectively.
Note that $\delta_{T} = \delta_{r} / 4$ and $\theta_{T} = \theta_{r}$.

2) Einstein equations during radiation domination are given by~\cite{Ma:1995ey}
\begin{eqnarray}
\label{eq:einstein1}
&& k^{2} \eta - \frac{1}{2} \frac{\dot a}{a} {\dot h} = - \frac{3}{2} \left( \frac{\dot a}{a} \right)^{2} \delta_{r} \,, \\
\label{eq:einstein2}
&& k^{2} {\dot \eta} = \frac{3}{2} \left( \frac{\dot a}{a} \right)^{2} ( 1 + w_{r} ) \theta_{r} \,, \\
\label{eq:einstein3}
&& {\ddot h} + 2 \frac{\dot a}{a} {\dot h} - 2 k^{2} \eta = - 3 \left( \frac{\dot a}{a} \right)^{2} c_{s, r}^{2} \delta_{r} \,, \\
\label{eq:einstein4}
&& {\ddot h} + 6 {\ddot \eta} + 2 \frac{\dot a}{a} ( {\dot h} + 6 {\dot \eta} ) - 2 k^{2} \eta \notag \\
&& = - 9 \left( \frac{\dot a}{a} \right)^{2} ( 1 + w_{r} ) \sigma_{r} \,,
\end{eqnarray}
where $w_{r} = 1/3$ and $c_{s, r}^{2} = \delta P_{r} / \delta \rho_{r} = 1 / 3$ are the equation of state and the sound speed of radiation, respectively.

Second, we discuss the initial conditions.
In the tight-coupling limit, adiabatic perturbations are given by
\begin{eqnarray}
\label{eq:tightcouple1}
&& \eta = 4 C \frac{1 - \cos x}{x^{2}} \,, \\
\label{eq:tightcouple2}
&& \dot h = - \frac{24C}{\tau} \left( \frac{2}{x} \sin x + \frac{2}{x^{2}} \cos x - \frac{2}{x^{2}} - 1 \right) \,, \\
\label{eq:tightcouple3}
&& \delta_{r} = - 8 C \left(\frac{2}{x} \sin x - \cos x + \frac{2}{x^{2}} \cos x - \frac{2}{x^{2}} \right) \,, \\
\label{eq:tightcouple4}
&& \theta_{\chi^0} = \theta_{r} = - \frac{6 C}{\tau} \left(- x \sin x - 2 \cos x + 2 \right) \,, \\
\label{eq:tightcouple5}
&& \delta_{\chi^0} = 12C \left( \frac{1 - \cos x}{x^{2}} + \frac{1}{2} \cos x - \frac{\sin x}{x} \right) \,,
\end{eqnarray}
where $x = k \tau / \sqrt{3}$ and $C$ is the initial amplitude of $\eta/2$.
This set of initial conditions fixes the residual gauge degrees of freedom in the synchronous gauge~\cite{Kamada:2016qjo}.
For comparison, cold dark matter ($\theta_{\rm cdm} = 0$) evolves as
\begin{eqnarray}
\delta_{\rm cdm} &=& 12 C \left( \frac{1 - \cos x}{x^{2}} + {\rm Ci}(x) - \ln x - \frac{\sin x}{x} \right) \,,
\end{eqnarray}
with the cosine integral Ci.

\begin{figure}[!h]
\centering
\includegraphics[scale=0.6]{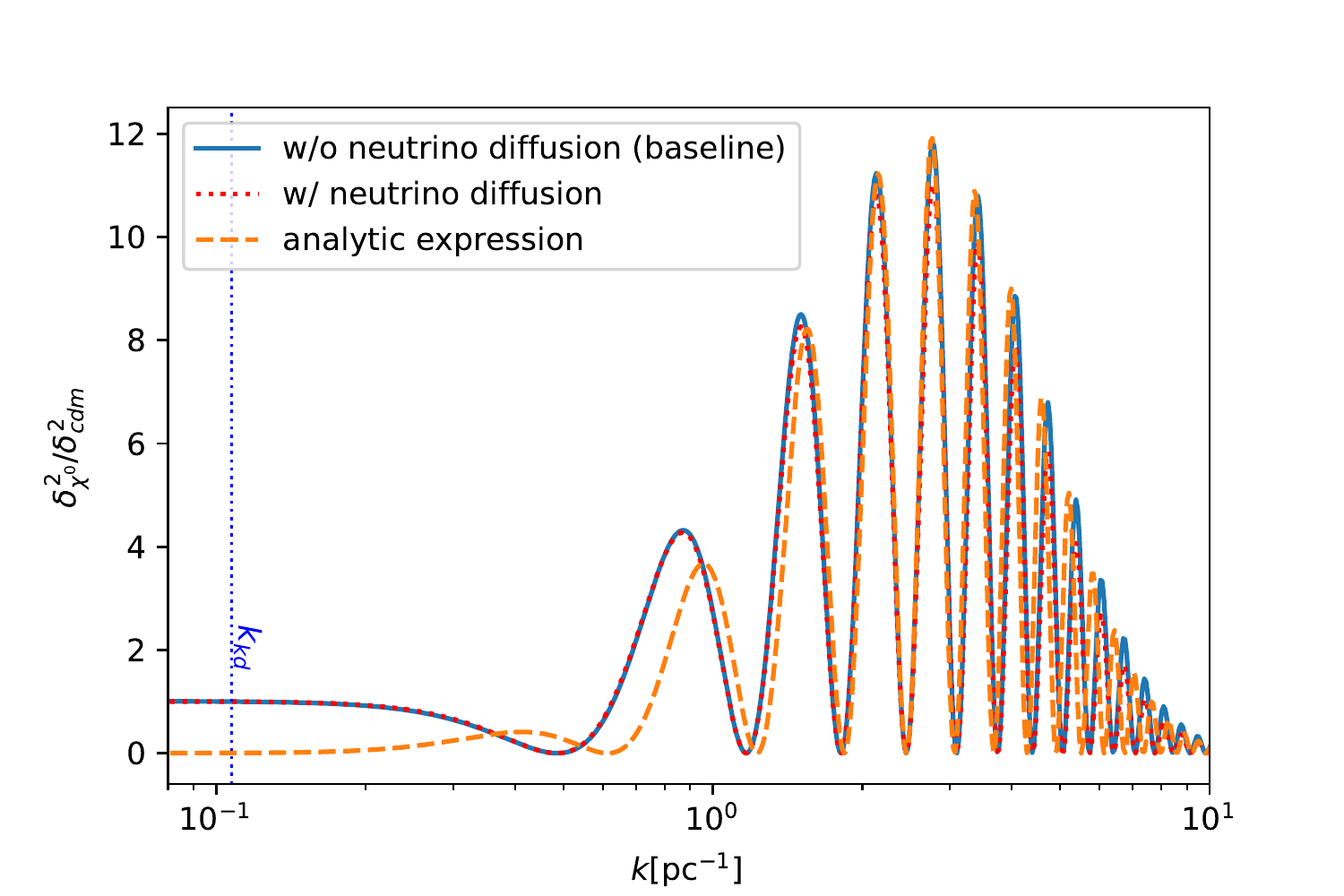}
\caption{Comparison of the power spectrum ratios at $\tau_{\rm eq}$ with and without neutrino diffusion.
The analytic expression given by Eqs.~\eqref{eq:fit1} and \eqref{eq:fit2} is also shown.}
\label{fig:nudiff}
\end{figure}

Third, we compare our numerically obtained power spectrum ratio (Fig.~\ref{fig:ratio} in the main text) with an analytic result.
References~\cite{Kamada:2017oxi} developed an analytic approach to the evolution of the density perturbations by assuming $\Gamma / H = (\tau_{\rm kd} / \tau)^{n}$.
One finds that the resultant power spectrum ratio is approximated by
\begin{eqnarray} 
&& \frac{\delta_{\chi^{0}}^{2}}{\delta_{\rm cdm}^{2}} \approx c_{\cal N} \frac{4 \pi}{n} \exp \left(- \frac{k}{k_{\rm damp}} \right) \left( \frac{k}{2\sqrt{3}k_{\rm kd}} \right)^{3} \notag \\
&& \quad \quad \quad ~ \times \sin^{2} \left( \frac{k}{\sqrt{3} k_{\rm kd}} \right) \,, \label{eq:fit1}\\
&& k_{\rm damp} = c_{\rm damp} \frac{n}{\pi} \sqrt{3}k_{\rm kd} \,, \quad k_{\rm kd} = \frac{c_{\rm kd}}{\tau_{\rm kd}} \,,\label{eq:fit2}
\end{eqnarray}
for $k \gg \tau_{\rm kd}$ and $n \gg 1$.
All the fudge factors, $c_{{\cal N}}$, $c_{\rm damp}$, and $c_{\rm kd}$, are unity in the case of $\Gamma / H = (\tau_{\rm kd} / \tau)^{n}$.
For our wino dark matter, we numerically find $n \simeq 15$ at $\tau = \tau_{\rm kd}$.
The above expression matches to the numerical result, when the fudge factors are $c_{{\cal N}} \simeq 0.91$, $c_{\rm damp} \simeq 0.95$, and $c_{\rm kd} \simeq 1.05$, as in Fig. \ref{fig:nudiff}.

Fourth, we discuss the impact of neutrino diffusion on the resultant power spectrum ratio.
The neutrino starts to free-stream around $T \sim 2 \MeV$.
One needs to derive and follow the neutrino full Boltzmann hierarchy with collision terms to take into account diffusion consistently.
Since it is beyond the scope of this work, here we take a practical approach instead.
We introduce the collision term by hand as
\begin{eqnarray}
\label{eq:cont_gamma}
&& {\dot \delta}_{\gamma} + \frac{4}{3} \theta_{\gamma} + \frac{2}{3} {\dot h} = a \frac{r_\nu}{r_\gamma} \gamma_{\nu0} (\delta_{\nu} - \delta_{\gamma})
 \,, \\
\label{eq:eular_gamma}
&& {\dot \theta}_{\gamma} - \frac{1}{4} \delta_{\gamma} = a \frac{r_\nu}{r_\gamma} \gamma_{\nu1} (\theta_{\nu} - \theta_{\gamma}) \,, \\
\label{eq:sigma_gamma}
&& \sigma_{\gamma} = 0 \,,
\end{eqnarray}
and
\begin{eqnarray}
\label{eq:cont_nu}
&& {\dot \delta}_{\nu} + \frac{4}{3} \theta_{\nu} + \frac{2}{3} {\dot h} = a \gamma_{\nu0} (\delta_{\gamma} - \delta_{\nu}) \,, \\
\label{eq:eular_nu}
&& {\dot \theta}_{\nu} - \frac{1}{4} \delta_{\nu} =  a \gamma_{\nu 1} (\theta_{\gamma} - \theta_{\nu}) \,, \\
\label{eq:ell2_nu}
&& {\dot \sigma}_{\nu} - \frac{4}{15} \theta_{\nu} + \frac{3}{10} k F_{\nu 3} - \frac{2}{15} {\dot h} - \frac{4}{5} {\dot \eta}  = - a \gamma_{\nu 2} \sigma_{\nu} \,,  \\
\label{eq:ell3_nu}
&& {\dot F}_{\nu,\ell} - \frac{k}{2 \ell - 1} \left[ \ell F_{\nu, \ell} - (\ell + 1) F_{\nu, \ell + 1} \right]  = - a \gamma_{\nu\ell} F_{\nu, \ell} \notag \\
&& \quad \quad \quad \quad \quad \quad \quad \quad \quad \quad \quad \quad \quad \quad \quad {\rm for} ~ \ell \ge 3 \,, 
\end{eqnarray}
where we followed the notation of Ref.~\cite{Ma:1995ey} except for the collision terms.
Here, $\gamma_{\nu \ell}$ is the contribution of the neutrino collision term to the multipole $\ell$.
We approximate $\gamma_{\nu \ell}$ by
\begin{eqnarray}
\gamma_{\nu \ell} = (G_{F} T)^{2}  \left[c_{\nu e \ell} n_{e} (T) +c_{\nu \nu \ell} n_{\nu}(T) \right] \,,
\end{eqnarray}
where $c_{\nu e \ell}$ and $c_{\nu \nu \ell}$ are fudge factors.
The number densities of the electron and neutrino are $n_{e} (T) = n_{\nu} (T) = (3 / 2) (\zeta(3) / \pi^{2}) T^{3}$, counting 2 degrees of freedom for each.
We take both $c_{\nu e \ell}$ and $c_{\nu \nu \ell}$ to be unity in our numerical analysis, so that $\gamma_{\nu\ell}\equiv\gamma_{\nu}$ is independent of $\ell$.
$r_{\gamma} = 1 - r_{\nu}$ is the photon+electron fraction of radiation and
\begin{eqnarray}
r_{\nu} \simeq
\begin{cases}
0.405 & \text{after $e^{\pm}$ annihilation} \\
0.488 & \text{before $e^{\pm}$ annihilation}
\end{cases}
\end{eqnarray}
is the neutrino fraction of radiation.
We set $r_{\nu} = 0.488$ in our numerical analysis.

Meanwhile, in the wino equations [Eqs.~\eqref{eq:deltachi} and \eqref{eq:thetachi}], we set $\delta_{T} = \delta_{\gamma} / 4$ and $\theta_{T} = \theta_{\gamma}$.
In the Einstein equations [Eqs.~\eqref{eq:einstein1}--\eqref{eq:einstein4}], we set $\delta_{r} = r_{\gamma} \delta_{\gamma} + r_{\nu} \delta_{\nu}$, $\theta_{r} = r_{\gamma} \theta_{\gamma} + r_{\nu} \theta_{\nu}$, and $\sigma_{r} = r_{\gamma} \sigma_{\gamma} + r_{\nu} \sigma_{\nu}$.
Figure~\ref{fig:nudiff} compares the power spectrum ratios $\delta_{\chi^0}^2/\delta_{\rm cdm}^2$ at $\tau_{\rm eq}$ calculated with and without neutrino diffusion.
We find that neutrino diffusion affects the ratios only slightly.

Last, we comment on a caveat when one evaluates the present wino power spectrum.
We follow the evolution of the wino or cold dark matter density contrasts until the neutral wino kinetically decouples and $\delta_{\chi}$ starts to logarithmically grow in deep subhorizon as $\delta_{c}$ and extrapolate the growth till the matter radiation equality $\tau_{\rm eq}$.
In the calculation of the boost factor, we simply multiply the power spectrum ratio at $\tau_{\rm eq}$ to the present matter power spectrum in cold dark matter that is generated by a public code \texttt{CAMB}~\cite{Lewis:1999bs}.
We remark that a public code like \texttt{CAMB} or \texttt{CLASS}~\cite{Lesgourgues:2011re} takes into account only physics around last scattering.
Thus the above procedure ignores the effects of quantum chromodynamics (QCD), neutrino decoupling, and $e^{\pm}$ annihilation.
To be consistent, we should use just the free-streaming neutrino in cold dark matter but also take into account these effects in wino dark matter, when calculating $\delta_{\chi^{0}}^{2} / \delta_{\rm cdm}^{2}$.
However, it is beyond the scope of this work. 
In the following, instead, we estimate the impacts of these effects.

QCD crossover takes place around $T_{c} \simeq 180 \MeV$~\cite{Watanabe:2006qe, Saikawa:2018rcs} ($150\MeV$~\cite{Aoki:2009sc, Bazavov:2011nk}).
It could change, for example, the equation of state of primordial plasma up to 30\% and thus the evolution of the density contrast.
Our dark matter power spectrum would suffer from this uncertainty above $k \sim 2 \unit{/ pc}$, which is comparable with $k_{\rm fs}$.
If QCD phase transition is first order, the evolution of the density contrast would be further amplified~\cite{Schmid:1998mx}.

Although the power spectrum ratios do not change with and without neutrino decoupling (see Fig.~\ref{fig:nudiff}), the power spectrum itself is affected by neutrino decoupling even in cold dark matter.
In addition, $e^{\pm}$ annihilation below $T \simeq 511 \keV$ changes the equation of state of primordial plasma up to 10\%, which results in up to 10\% change in the density contrast with $k \sim 2 \times 10^{-3} \unit{/ pc}$~\cite{Bertschinger:2006nq}.

\begin{figure}[!h]
\centering
\includegraphics[scale=0.6]{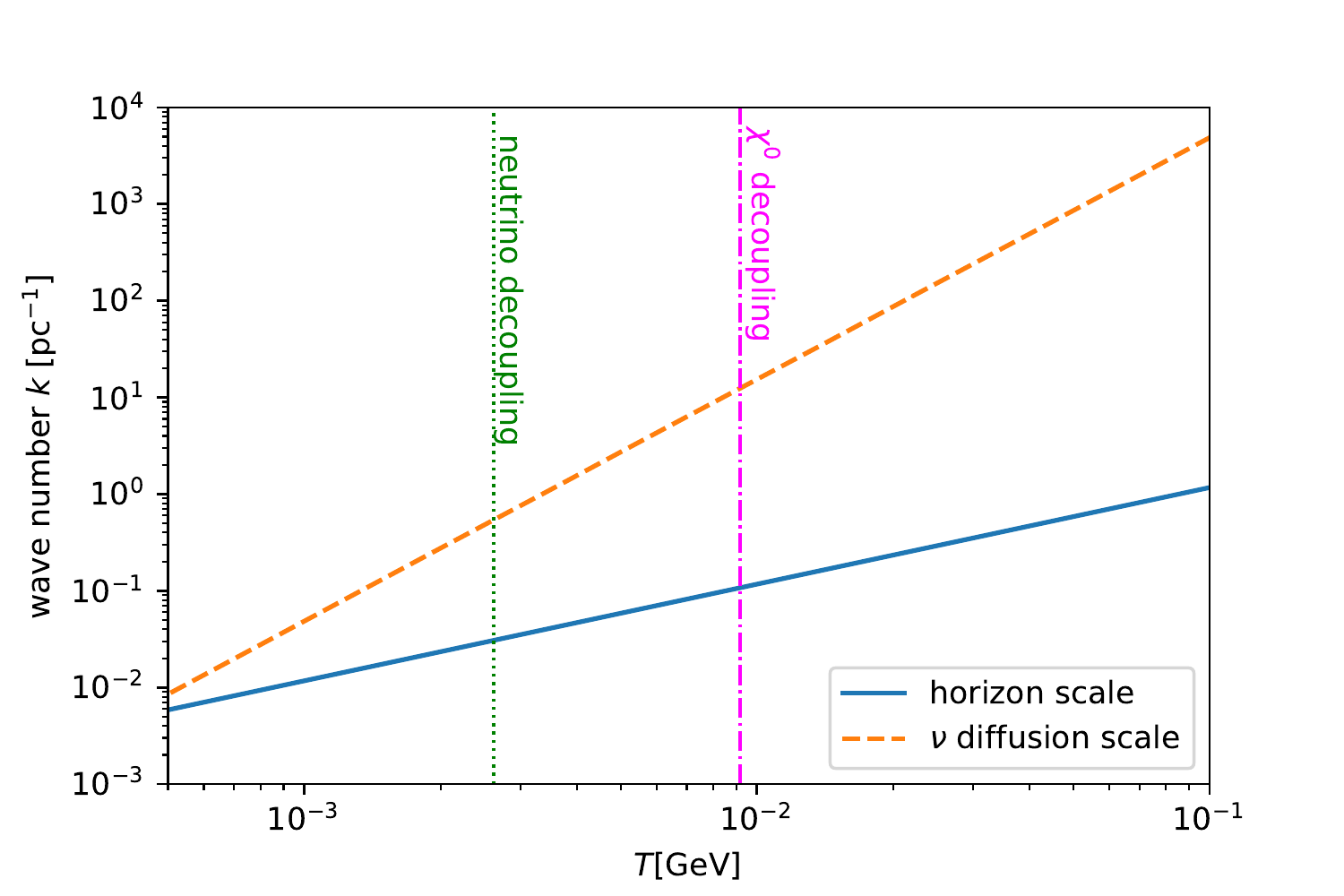}
\caption{Time evolution of the horizon and the neutrino diffusion scales.}
\label{fig:scales}
\end{figure}

\begin{figure}[!h]
\centering
\includegraphics[scale=0.6]{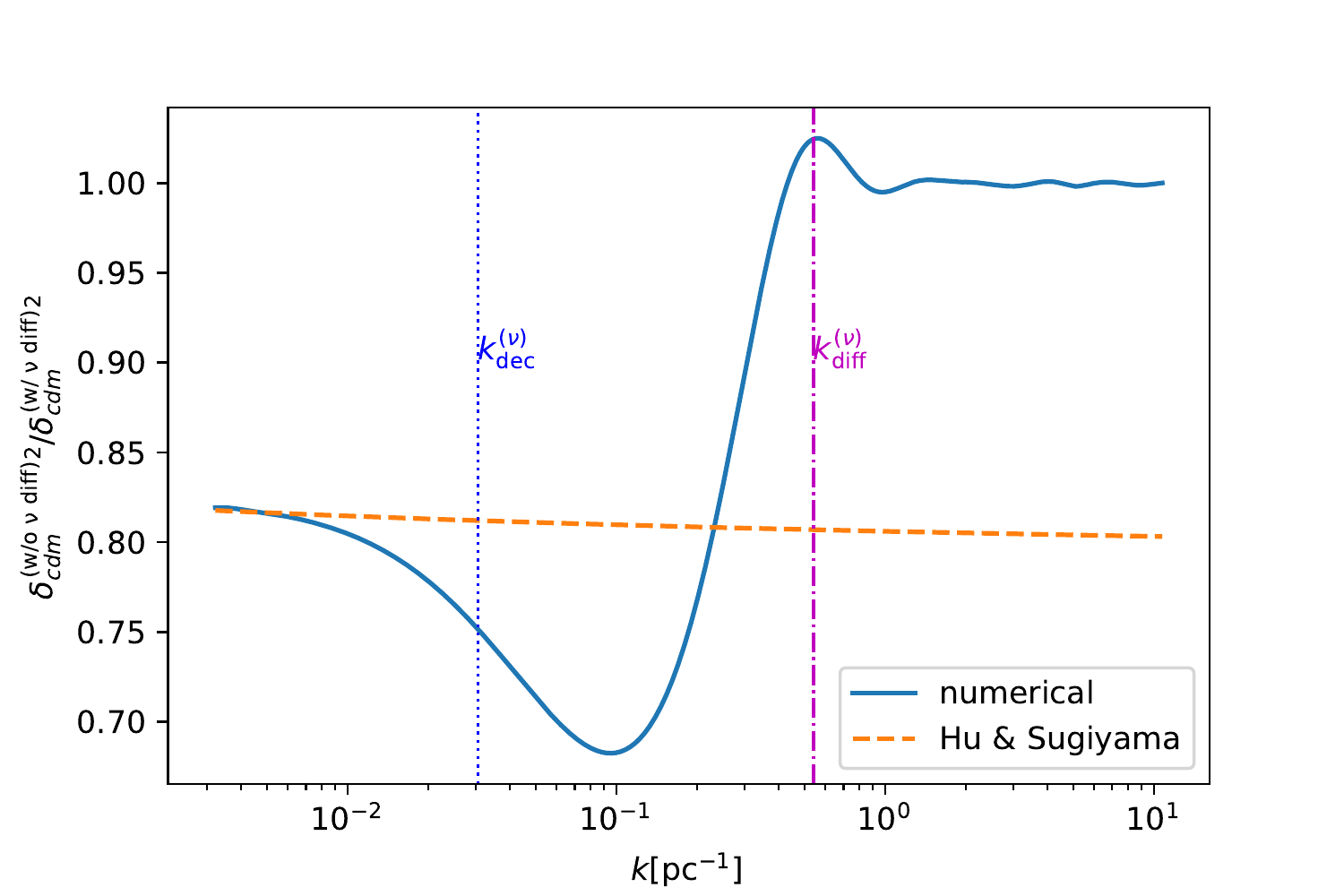}
\caption{Ratio of the cold dark matter power spectrum without neutrino diffusion to the one with diffusion.
Vertical lines are the horizon scale $k^{(\nu)}_{\rm dec}$ and the neutrino diffusion one $k^{(\nu)}_{\rm diff}$ at neutrino decoupling.}
\label{fig:cdm}
\end{figure}

Even large-scale modes, which are superhorizon at the time of neutrino decoupling, get enhanced by neutrino free-streaming when compared to the case of the interacting (perfect fluid) neutrino~\cite{Hu:1995en}.
We define the neutrino decoupling time $\tau^{(\nu)}_{\rm dec}$ as the time that satisfies $\gamma_{\nu} / H = 4$ in a similar manner to the kinetic decoupling of wino dark matter.
The diffusion scale $k^{(\nu)}_{\rm diff}$ can be estimated as~\cite{Jeong:2014gna}
\begin{equation}
{k^{(\nu)}_{\rm diff}(\tau)}^{-2} = \int d\tau \frac{8}{45} \frac{1}{\rho_{r} + P_{r}} \rho_{\nu} \tau_{\nu} \,,
\end{equation}
where $\tau_\nu= 1/ (a\gamma_\nu)$ is the conformal mean free time of the neutrino.
Figure~\ref{fig:scales} shows the evolution of the diffusion and horizon scales as function of the temperature.
To see the impact of free-streaming explicitly, in Fig.~\ref{fig:cdm} we show the ratio of the cold dark matter power spectra at $\tau_{\rm eq}$ with and without neutrino diffusion and free-streaming.
The change in the power spectrum is up to 15\%.
For reference, the horizon scale $ k^{(\nu)}_{\rm dec} (= 1/\tau^{(\nu)}_{\rm dec})$ and the neutrino diffusion scale $k^{(\nu)}_{\rm diff} (= 1/\tau^{(\nu)}_{\rm diff})$ at the neutrino decoupling are also shown.
One can see a transition in the ratio, which takes place at $k^{(\nu)}_{\rm dec}\lesssim k\lesssim k^{(\nu)}_{\rm diff}$.
Above this neutrino diffusion hardly affects the cold dark matter perturbations.
Below this the ratio converges to the result of Ref.~\cite{Hu:1995en} that estimates the effect of neutrino free-streaming on cold dark matter perturbations.

\section{Calculation of Boost Factor}
\label{sec:calcB}
We briefly summarize how we estimate the boost factor according to Ref.~\cite{Hiroshima:2018kfv}.
To compute Eq.~\eqref{eq:boost} in the main text, one is required to know the mass function and inner density profile of subhalos, which can be achieved by tracing how subhalos form and evolve.
Evolution of subhalos can be divided into two stages with accretion onto host halos as a border.

Before accretion, subhalos form and evolve as field halos.
With their evolution well described by the extended Press-Schechter formalism, their accretion rate onto a progenitor of the host can be also given accordingly.
Denoting the subhalo mass and the redshift at accretion respectively as $m_{\rm acc}$ and $z_{\rm acc}$,
the accretion rate ${d^2N_{\rm sh}}/{d\ln m_{\rm acc}/dz_{\rm acc}}$ we adopt is~\cite{Yang:2011rf}
\begin{eqnarray}
\label{eq:accrate}
\frac{d^{2} N_{\rm sh}}{d\ln m_{\rm acc} dz_{\rm acc}} = {\cal F} \left( s_{\rm acc},\delta_{\rm acc} | S_{0}, \delta_{0}; {\bar M}_{\rm acc} \right) \frac{ds_{\rm acc}}{dm_{\rm acc}} \frac{d{\bar M}_{\rm acc}}{dz_{\rm acc}}
\,, \notag \\
\end{eqnarray}
with ${\bar M}_{\rm acc}$ being the (mean) mass of a host progenitor (that eventually evolves into a mass $M_{0}$ at $z = z_{0}$) at $z_{\rm acc}$.
In Eq.~\eqref{eq:accrate}, $s_{\rm acc} = \sigma^{2} (m_{\rm acc}, z=0)$ is the variance of overdensity smoothed at a scale corresponding to $m_{\rm acc}$ and
$\delta_{\rm acc} = \delta_c (z_{\rm acc}) =1.686 / D (z_{\rm acc})$ is the critical overdensity at $z_{\rm acc}$. Similarly, $S_{0} = \sigma^{2} (M_{0}, z=0)$ and $\delta_{0} = \delta_{\rm c} (z)$, which gives the boundary condition for the host halo evolution.
The definition of ${\cal F}$ in Eq.~\eqref{eq:accrate} is given shortly below.

For the mass evolution of the host, we adopt the fitting formula given in Ref.~\cite{Correa:2014xma}.
The probability distribution of the host mass $P \left( M_{\rm acc} | S_{0}, \delta_{0} \right)$ approximately follows the log-normal distribution with a logarithmic dispersion of
\begin{eqnarray}
\sigma_{\log M_{\rm acc}} = 0.12 - 0.15 \log \left( \frac{{\bar M}_{\rm acc}}{M_{0}} \right).
\end{eqnarray}
The mean value of ${\bar M}_{\rm acc}$ is given by
\begin{eqnarray}
&& {\bar M}_{\rm acc} \left( z | M_{0}; z=0 \right) = M_{0} (1+z)^{\alpha} \exp(\beta z) \,, \\
&& \beta = - g(M_{0}) \,,\\
&& \alpha = \left[ \frac{1.686 \sqrt{2 / \pi}}{D^{2} (z=0)} \left. \frac{dD}{dz} \right|_{z=0} + 1 \right] g(M_{0}) \,, \\
&& g(M_{0}) = \left[ S_{0} (M_{0} / q) - S_{0} (M_{0}) \right]^{-1/2} \,, \\
&& q = 4.137 \tilde{z}_{f}^{-0.9476} \,,\\
&& \tilde{z}_{f} = -0.0064 (\log M_{0})^{2} + 0.0237 (\log M_{0}) +1.8837 \,. \notag \\
\end{eqnarray}

The definition of ${\cal F}$ in Eq.~\eqref{eq:accrate} is given as 
\begin{eqnarray}
&& {\cal F} \left( s_{\rm acc}, \delta_{\rm acc} | S_{0}, \delta_{0}; {\bar M}_{\rm acc} \right) \notag \\ 
&& = \int \Phi(s_{\rm acc}, \delta_{\rm acc} | S_{0}, \delta_{0}; M_{\rm acc}) P \left( M_{\rm acc} | S_{0}, \delta_{0} \right) dM_{\rm acc} \,, \notag \\ \\
&& \Phi \left( s_{\rm acc}, \delta_{\rm acc} | S_{0}, \delta_{0}; M_{\rm acc} \right) \notag \\
&& = \left[ \int^{\infty}_{S(m_{\rm max})} ds_{\rm acc} F \left(s_{\rm acc}, \delta_{\rm acc} | S_{0}, \delta_{0}; M_{\rm acc} \right) ds_{\rm acc} \right]^{-1} \notag \\
&& \times
\begin{cases}
F \left( s_{\rm acc}, \delta_{\rm acc} | S_{0}, \delta_{0}; M_{\rm acc} \right) & (m_{\rm acc} \le m_{\rm max}) \\
0 &({\rm otherwise})
\end{cases} \,, \\
&& F \left( s_{\rm acc}, \delta_{\rm acc} | S_{0}, \delta_{0}; M_{\rm acc} \right) \notag \\
&& = \frac{1}{\sqrt{2 \pi}} \frac{\delta_{\rm acc} - \delta_M}{(s_{\rm acc} - S_{M})^{3/2}} \exp \left[ - \frac{\delta_{\rm acc} - \delta_{M}}{2 (s_{\rm acc} - S_{M})} \right]\, ,
\end{eqnarray}
where $m_{\rm max} = \min[M_{\rm acc}, M_{0} / 2]$, $M_{\rm max} = \min[ M_{\rm acc} + m_{\rm acc}, M_{0}]$, $S_{M} = \sigma^{2} (M_{\rm max})$, and $\delta_{M}$ is defined as $\delta_{\rm c} (z)$ at $z$ when $M = M_{\rm max}$.

In addition, as born as field halos, the inner density profile of subhalos at the moment of accretion 
is given by the Navarro-Frenk-White (NFW) one~\cite{Navarro:1995iw, Navarro:1996gj},
\begin{eqnarray}
\rho = \frac{\rho_{s}}{(r / r_{s}) (1 + r / r_{s})^{2}} \,.
\end{eqnarray}
The concentration parameter of subhalos is given by $c = r_{v} / r_{s}$, where $r_{v}$ is their virial radius.
We assume that the concentration parameter obeys the log-normal distribution with the standard deviation of 
$\sigma_{\log c} = 0.13$~\cite{Ishiyama:2011af}.
For the mean value of the halo concentration, we adopt the fitting formula in Ref.~\cite{Correa:2015dva}. 
In terms of $c_{200} = c \, r_{200} / r_{v}$ (see Ref.~\cite{Hu:2002we} for conversion), with $r_{200}$ being the radius within which the averaged mass density is 200 times as large as the homogeneous one, the concentration-mass relation is given as 
\begin{eqnarray}
&& \log c_{200} = \alpha + \beta \log \left( \frac{M_{\rm 200}}{M_{\odot}} \right) \left[ 1 + \gamma \log^{2} \left( \frac{M_{\rm 200}}{M_{\odot}} \right) \right] \,, \notag \\ \\
&& \alpha = 1.7543 - 0.2766 (1+z) + 0.02039 (1+z)^{2} \,,\\
&& \beta = 0.2753 - 0.00351 (1+z) - 0.3038 (1+z)^{0.0269} \,,\notag \\ \\
&& \gamma = - 0.01537 - 0.02102 (1+z)^{-0.1475} \,,
\end{eqnarray}
for $z\le4$ and 
\begin{eqnarray}
&& \log c_{200} = \alpha + \beta \log \left( \frac{M_{\rm 200}}{M_{\odot}} \right) \,,\\
&& \alpha = 1.3081 - 0.1087 (1+z) + 0.00398 (1+z)^{2} \,, \notag \\ \\
&& \beta = 0.0223 - 0.0944 (1+z)^{-0.3907}\,,
\end{eqnarray}
for $z>4$.
In contrast to the concentration-mass relation extrapolated from galaxy- or cluster-sized halos with a single power law (for instance, Ref.~\cite{Neto:2007vq}), Ref.~\cite{Correa:2015dva} gives one flattened towards smaller masses.

Once accreted onto host halos, subhalos undergo tidal mass splitting as they orbit in the gravitational field of hosts.
The mass-loss rate we adopt is~\cite{Hiroshima:2018kfv, Ando:2019xlm}
\begin{eqnarray}\label{eq:massloss}
\dot m(z) = - A \frac{m(z)}{\tau_{\rm dyn}(z)} \left[ \frac{m(z)}{M(z)} \right]^{\zeta} \,,
\end{eqnarray}
where 
\begin{eqnarray}
\tau_{\rm dyn} = \sqrt{\frac{3 \pi}{16 G {\bar \rho}_{h}}}
\end{eqnarray}
is the dynamical time scale associated to the host halo~\cite{Jiang:2014nsa}, whose mean density is ${\bar \rho}_{h}$, and the coefficients $A$ and $\zeta$ are given as 
\begin{eqnarray}
&& \log A = \left[- 0.0003 \log \left( \frac{M (z)}{M_{\odot}} \right) + 0.02 \right] z \notag \\
&& \quad \quad \quad ~ + 0.011 \log \left( \frac{M (z)}{M_{\odot}} \right) - 0.354 \,, \\
&& \zeta = \left[0.00012 \log \left( \frac{M (z)}{M_{\odot}} \right) - 0.0033\right] z \notag \\
&& \quad \quad - 0.0011 \log \left( \frac{M (z)}{M_{\odot}} \right) + 0.026 \,.
\end{eqnarray}
$G$ is the gravitational constant.
The mass-loss rate is integrated to give mass evolution of subhalos after accretion.
Given mass evolution, subhalo mass function can be obtained from the initial condition given by the accretion rate Eq.~\eqref{eq:accrate}.
In addition, we assume that the subhalo profile is given as the NFW one truncated at $r_t$,
\begin{eqnarray}
\rho = \begin{cases}
\label{eq:tNFW}
\dfrac{\rho_{s}}{(r / r_{s}) (1 + r / r_{s})^{2}} & (r\le r_t) \\
0 & ({\rm otherwise}) 
\end{cases}
\,.
\end{eqnarray}
We assume that $\rho_{s}$ and $r_{s}$ evolve according to Ref.~\cite{Penarrubia:2010jk} in the course of tidal stripping, and then obtain evolution of $r_{t}$ from the mass-conservation condition~\cite{Hiroshima:2018kfv, Ando:2019xlm}.

Provided the above prescription for the evolution of subhalos, one can solve Eq.~\eqref{eq:boost} in the main text numerically.
Specifically, we compute 
\begin{eqnarray}
&& B(M) = \frac{1}{{\bar L}(M)} \int d\ln m_{\rm acc} \int dz_{\rm acc} \frac{dN_{\rm sh}}{d\ln m_{\rm acc} dz_{\rm acc}} \notag \\
&& \quad \quad \quad \quad \times \int dc_{\rm acc} P \left( c_{\rm acc} | m_{\rm acc} , z_{\rm acc} \right) L_{\rm sh} \Theta[r_{t} - 0.77 r_{s}] \,, \notag \\
\end{eqnarray}
where $P\left( c_{\rm acc} | m_{\rm acc}, z_{\rm acc} \right)$ is the probability distribution of the $c_{\rm acc}$, which is given above. 
In the above expression, the Heaviside function indicates that subhalos whose truncation radius is smaller than 0.77 times $r_s$ do not contribute to the boost factor, since those subhalos are supposed to be totally disrupted according to Ref.~\cite{Hayashi:2002qv}.
Omitting substructures inside subhalos (i.e., subsubhalos), the luminosity of each subhalo $L_{\rm sh}$ is proportional to a volume integral of the density squared.
The profile in Eq.~\eqref{eq:tNFW} yields
\begin{eqnarray}
\label{eq:Lsh}
L_{\rm sh} \propto \rho_{s}^{2} r_{s}^{3} \left[ 1 - \frac{1}{(1 + r_{t} / r_{s})^{3}} \right],
\end{eqnarray}
which is specified once $m_{\rm acc}$, $z_{\rm acc}$, and the concentration parameter at the accretion
$c_{\rm acc}$ are given.
It is rather straightforward to incorporate effects of subsubhalos (and even their substructures)~\cite{Hiroshima:2018kfv, Ando:2019xlm}.
Meanwhile, the luminosity of the host's smooth component ${\bar L}(M)$ can be obtained by replacing 
$r_{t} / r_{s}$ with the concentration parameter of the host $c$ in Eq.~\eqref{eq:Lsh}.

\begin{figure}[!h]
\centering
\includegraphics[scale=0.45]{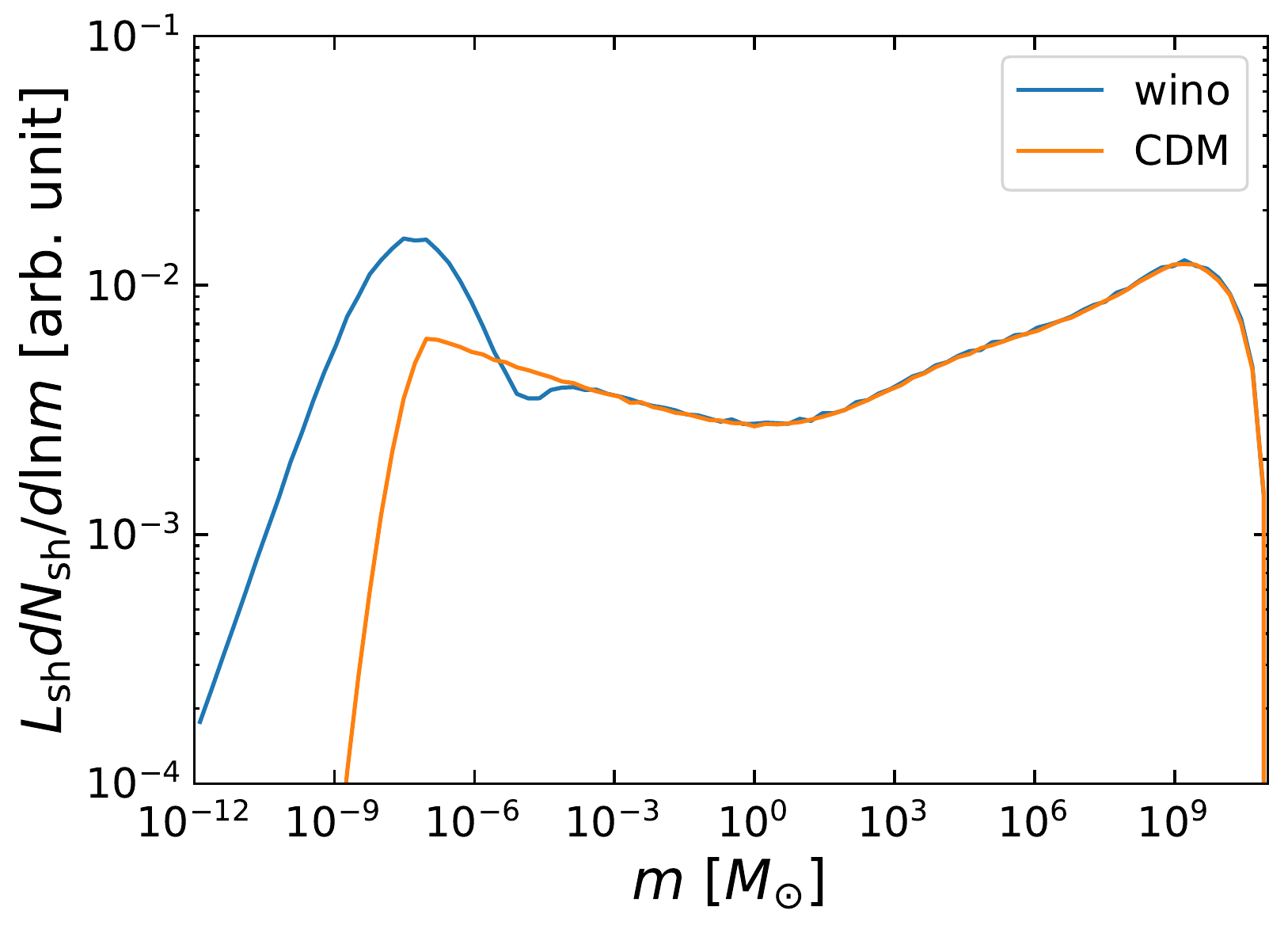}
\caption{The subhalo mass function $dN_{\rm sh} / d\ln m$ multiplied by the gamma-ray luminosity from dark matter annihilation in the subhalos $L_{\rm sh}$ (in arbitrary units) for a host halo with mass $M_{\rm host} =  10^{12} M_{\odot}$ in the case of wino dark matter (blue) and CDM (orange).}
\label{fig:mf}
\end{figure}

Figure~\ref{fig:mf} shows the integrand of Eq.~\eqref{eq:boost} in the main text, $L_{\rm sh} dN_{\rm sh} / d\ln m$, for a host halo with the mass of $M_{\rm host} = 10^{12} M_{\odot}$. 
It is manifested that the subhalo contributions to the boost factor are significantly enhanced around $m=M_{\rm fs}$ and suppressed at smaller scales.
This is exactly what we expect from the matter power spectrum in Fig.~\ref{fig:ratio} in the main text, which exhibits boosted acoustic peaks and suppressed power within the free-streaming length.
Even though this affects only subhalos with very small masses, $m \lesssim 10^{-5} M_{\odot}$, we find that the overall boost factor (after integrating over the subhalo masses) becomes larger than that of a na\"{i}ve model with a sudden cutoff at $M_{\rm fs} = 10^{-7} M_{\odot}$ (which we refer to as CDM) by about 30\%.

\bibliography{sdmhtw} 

\end{document}